\documentclass[conference]{IEEEtran}
\IEEEoverridecommandlockouts
\ifCLASSINFOpdf
\else
\fi
\usepackage[top=0.75in, bottom=0.75in, left=0.75in, right=0.75in]{geometry}
\usepackage{bbm}
\usepackage{verbatim}
\usepackage{graphicx}
\usepackage{cite}
\usepackage{url}
\usepackage[cmex10]{amsmath}
\usepackage{amssymb}
\usepackage{amsmath}
\usepackage{algorithm, algorithmicx, algpseudocode}
\usepackage[caption=false,font=footnotesize]{subfig}
\usepackage{color}
\usepackage{cite}
\usepackage{epstopdf}
\usepackage{calc}
\usepackage{array}

\DeclareMathOperator*{\argmin}{arg\,min} 

\newcommand{\Nr}[0]{N_{\text{R}}}

\newcommand{\hermitian}[0]{\text{H}}
\newcommand{\transpose}[0]{\text{T}}

\begin{document}
%
\title{DSP Linearization for Millimeter-Wave All-Digital Receiver Array with Low-Resolution ADCs}

\author{
\IEEEauthorblockN{Han Yan 
     and Danijela Cabric}
\IEEEauthorblockA{
Electrical and Computer Engineering Department, University of California, Los Angeles\\
Email: yhaddint@ucla.edu, danijela@ee.ucla.edu}
\thanks{This work was supported in part by NSF under grant 1718742.}
\thanks{This work was supported in part by the ComSenTer and CONIX Research Centers, two of six centers in JUMP, a Semiconductor Research Corporation (SRC) program sponsored by DARPA.}
}


\IEEEspecialpapernotice{(Invited Paper)
}


\maketitle

\begin{abstract}
Millimeter-wave (mmWave) communications and cell densification are the key techniques for the future evolution of cellular systems beyond 5G. Although the current mmWave radio designs are focused on hybrid digital and analog receiver array architectures, the fully digital architecture is an appealing option due to its flexibility and support for multi-user multiple-input multiple-output (MIMO). In order to achieve reasonable power consumption and hardware cost, the specifications of analog circuits are expected to be compromised, including the resolution of analog-to-digital converter (ADC) and the linearity of radio-frequency (RF) front end. Although the state-of-the-art studies focus on the ADC, the nonlinearity can also lead to severe system performance degradation when strong input signals introduce inter-modulation distortion (IMD). The impact of RF nonlinearity becomes more severe with densely deployed mmWave cells since signal sources closer to the receiver array are more likely to occur. In this work, we design and analyze the digital IMD compensation algorithm, and study the relaxation of the required linearity in the RF-chain. We propose novel algorithms that jointly process digitized samples to recover amplifier saturation, and relies on beam space operation which reduces the computational complexity as compared to per-antenna IMD compensation.
\end{abstract}


%
\IEEEpeerreviewmaketitle

\section{Introduction}
\label{sec:Introduction}
Millimeter-wave (mmWave) communications is a promising technology for future mobile networks due to abundant bandwidth. Currently, there is a total of 3.85 GHz mmWave spectrum allocated for 5G-NR, and it is anticipated more bandwidth will be open between mmWave to sub Terahertz (THz) band. As shown in both theory and measurements, the additional pathloss at higher frequencies can be handled by the increased array gain of an aperture with constant area due to smaller wavelength. In other words, large antenna array at both base station (BS) and user equipment (UE) will certainly be used. Although analog array and hybrid digital and analog array are the commonly considered architectures for 5G-NR mmWave receivers, the fully digital receiver array has great potential \cite{Zorzi_Rx_array,another_rx_array}. A key challenge in system design is to determine the hardware specification requirements for the radio frequency (RF) circuits. It is partly because the circuits designs become more challenging in mmWave/sub-THz bands. However, using reasonably compromised specifications helps reduce the overall power consumption.

One of the major system bottlenecks in digital array architecture is the analog to digital converter (ADC) because of excessive power consumption when high bandwidth and high precision are targeted. Fortunately, the information loss from the low resolution ADCs in large array regime can be effectively controlled by advanced signal processing  \cite{adc_survey,heath_overview}. In fact, maximum likelihood estimator provides adequate detection performance even with one bit quantization \cite{adc1bit_heath}. Trade-off between performance, baseband processing complexity, and ADC power are actively studied \cite{heath_adc_UL,studer_adc}. In addition, systems using conventional linear MIMO processing also benefit from increased antenna array size by reducing the distortion from ADCs \cite{bjorson_MIMO_HW_scaling}. Another key design specification is the transceiver linearity. Current research mostly focuses on the transmitter linearity problem. Emerging solutions include power amplifier design technology \cite{5gPA} and digital distortion mitigation. However, the impact of nonlinearity in large antenna array requires further studies. Also, there is a lack of investigation of the receiver side, except the work in \cite{UCSB_LNA_ADC,studer_impairment_asilomar}, where the impact of nonlinearity and ADC quantization error are analyzed. In summary, digital linearization techniques suitable for mmWave massive MIMO receivers are rarely investigated in the literature.

In this work, we focus on the digital signal processing (DSP) algorithm design for compensation of receiver nonlinearity in mmWave radios. We consider both small scale nonlinearity where distortion can be approximated by polynomial, and large scale nonlinearity where saturation occurs. The design is tailored for low resolution ADCs which are critical elements of future mmW fully digital architectures. 

The rest of the paper is organized as follows. In Section~\ref{sec:system_model}, we introduce the system model. Section~\ref{sec:algorithm} includes DSP algorithm for compensation of LNA nonlinear distortion in large scale mmWave receiver array. The efficacy of the proposed algorithm is numerically evaluated in Section~\ref{sec:simulation results}. Section~\ref{sec:Conclusion} concludes the paper.

\textit{Notations:} Scalars, vectors, and matrices are denoted by non-bold, bold lower-case, and bold upper-case letters, respectively, e.g. $h$, $\mathbf{h}$ and $\mathbf{H}$. The element in $i$-th row and $j$-th column in matrix $\mathbf{H}$ is denoted by $\{\mathbf{H}\}_{i,j}$. Transpose and Hermitian transpose are denoted by $(.)^{\transpose}$ and $(.)^{\hermitian}$, respectively. The $l_2$-norm of a vector $\mathbf{h}$ is denoted by $||\mathbf{h}||$. $\text{diag}(\mathbf{A})$ aligns diagonal elements of $\mathbf{A}$ into a vector, and $\text{diag}(\mathbf{a})$ aligns vector $\mathbf{a}$ into a diagonal matrix.

%
%
\section{System Model}
\label{sec:system_model}

We consider a single-cell uplink system where a fully digital mmWave/THz BS has $\Nr$ receiver antennas and RF-chains. In this network, there are $U$ users (UE) close to BS. Since we focus on the receiver side of the BS, we do not specify antenna size or array architecture of UEs. Focusing on high signal to noise ratio (SNR) regime, the received signal at $\Nr$ antennas, $\mathbf{x} = [x_1,\cdots,x_{N_{\text{R}}}]$, is denoted as
\begin{align}
\mathbf{x} = \mathbf{H}\mathbf{s}
\label{eq:basic_model}
\end{align}
where $\mathbf{x}\in\mathbb{C}^{\Nr}$ is the received sample vector at one time instance. $\mathbf{s}\in\mathbb{C}^{U}$ contains the transmit samples from the $U$ UEs. The $u$-th column of $\mathbf{H}$, $\mathbf{h}_u$, is the post-transmitter-precoding channel between $u$-th UE and $\Nr$ antennas at the BS. In other words, for each user the channel is effectively single input multiple output (SIMO). We assume that multi-user channel $\mathbf{H}$ is perfectly known to the BS. We also assume the SIMO channel is purely geometric, i.e., $[\mathbf{h}_k]_n = e^{j\pi(n-1)\sin(\theta_k)}$ where $\theta_k$ is the AoA of $k$-th stream. Different pathlosses are modeled by signal strength of each stream $\sigma_i^2 = \mathbb{E}(|s_i|^2)$, assumed to be known. 

We are particularly interested in a system where the LNA of BS is non-ideal. Using the baseband equivalent nonlinearity model \cite{UCSB_LNA_ADC}, the output of LNA $\mathbf{y}_{\text{LNA}} = [y_{\text{LNA},1},\cdots, y_{\text{LNA},n}]^{\transpose}$ is denoted as
\begin{align}
y_{\text{LNA},n} = p(x_n)=
\begin{cases}
\beta_1 x_n + \beta_3 x_n|x_n|^2+w_n, & |x_n| \leq V_{\text{sat}}\\
\frac{x_n}{|x_n|} V_{\text{max}}+w_n, & |x_n| > V_{\text{sat}}
\end{cases}
\label{eq:LNA_NL}
\end{align}
where $\beta_1$, $\beta_3$, $V_{\text{IP}3}$ are the nonlinearity parameters defined by the input corresponding 3rd order input interceptor point (IIP3)\footnote{Its specific value is determined by LNA linear gain as well as IIP3 , e.g., $\beta_1 = 56, \beta_3 = -7497, V_{\text{IP}3} = 0.05$ Volt for an LNA with 35dB linear gain and -10 dBm IIP3. The value $V_{\text{sat}} = V_{\text{IP}3}/2$ according to \cite{UCSB_LNA_ADC}. For clarity, we remove the unit Volt in $V_{\text{IP}3}$ in the rest of the paper.}. $V_{\text{max}}$ is the maximum output magnitude, i.e., $V_{\text{max}} = \beta_1V_{\text{sat}} + \beta_3 V^3_{\text{sat}}$. $w_n$ is the thermal noise, and it is omitted in this work since study focuses on strong signal regime. Note that the distortion term $x_n|x_n|^2$ leads to multiplication among data samples from different UEs, and is referred as inter-modulation distortion (IMD), and the saturation refers to scenario where the output magnitude is not dependent of input when $|x_n|>V_{\text{sat}}$.

We consider scenario where each RF-chain at BS uses low resolution ADC with $B$ bits of quantization for both real and imaginary parts after the down-conversion and automatic gain controller (AGC). The ADC output is given by
\begin{align}
y_{\text{ADC},n} = \mathcal{Q}_B(y_{\text{LNA},n})
\label{eq:ADC_quantization}
\end{align}
where $\mathcal{Q}_B(.)$ is the complex quantization function which consists of granular and overload regions \cite{UCSB_LNA_ADC}. 

We focus on the linear MIMO detection  
\begin{align}
\hat{\mathbf{s}} = \mathbf{W}^{\hermitian}\mathbf{y}_{\text{ADC}}
\end{align}
where the MIMO combiner is based on zero-forcing (ZF) criterion, i.e., $\mathbf{W}^\hermitian = (\mathbf{H}^{\hermitian}\mathbf{H})^{-1}\mathbf{H}^{\hermitian}$.

\begin{figure}
\begin{center}
\includegraphics[width=0.5\textwidth]{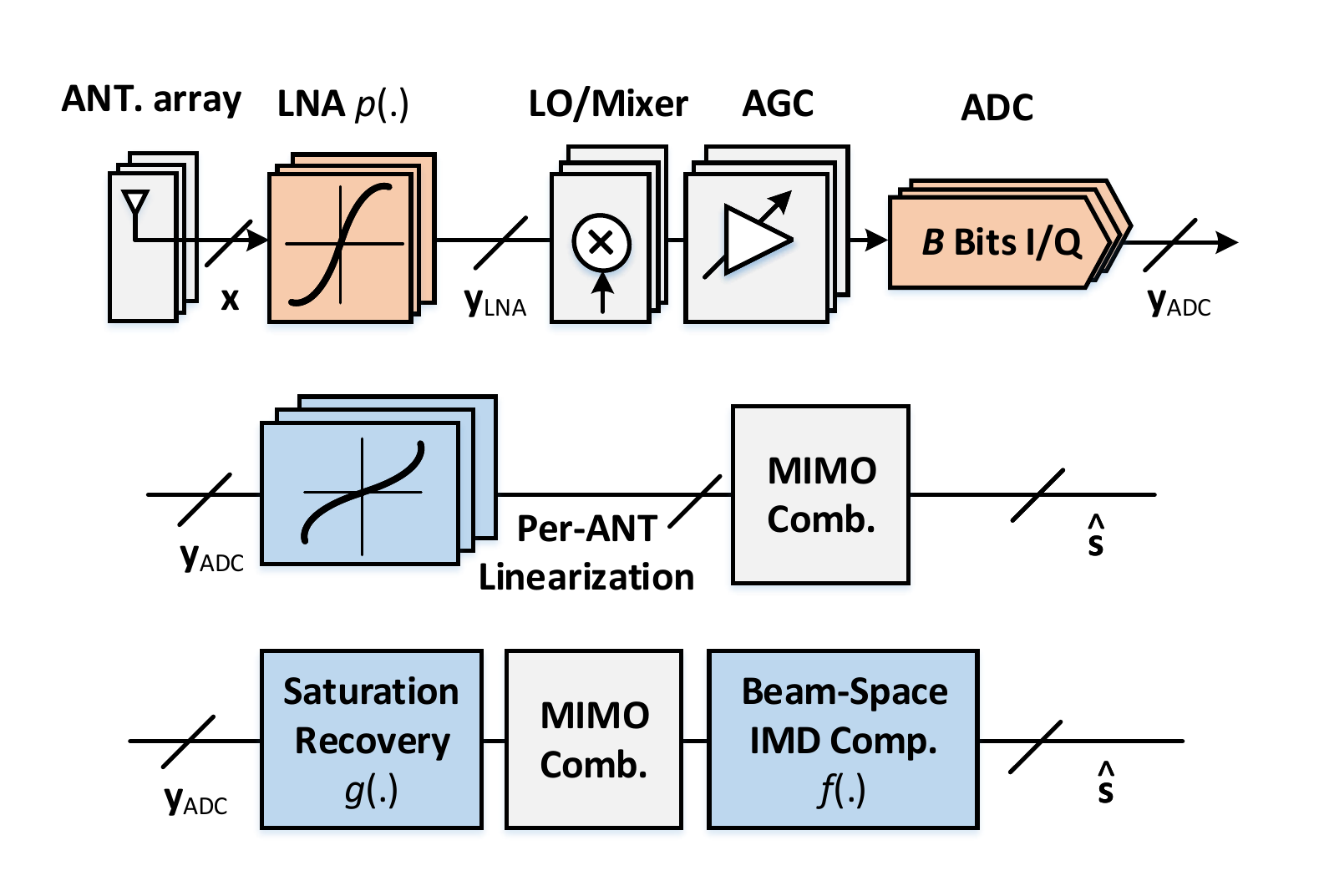}
\end{center}
\vspace{-2mm}
\caption{System model of mmWave UL with fully digital receiver arrays. First row focuses on the hardware impairments we study in this work. Second row is existing solution for per-antenna linearization. Third row is the proposed framework.}
\vspace{-2mm}
\label{fig:system_model}
\end{figure}

Due to LNA nonlinearity distortion and ADC quantization, the mean square error $\text{MSE} = \mathbb{E}|\mathbf{s}-\hat{\mathbf{s}}|^2$ and the corresponding symbol decoding suffer from additional error. In order to mitigate this error, DSP linearization algorithms can be used. These approaches use a known small scale nonlinearity model to invert (\ref{eq:LNA_NL}). Conventional receiver linearization is tailored for single RF-chain, and a straightforward extension for digital array is to use this approach for each LNA, as shown in Figure~\ref{fig:system_model} second row. However, these algorithms would have prohibitive complexity for large antenna array size $\Nr$. Moreover, they cannot compensate impact of large scale nonlinearity.

Based on the receiver model in Figure~\ref{fig:system_model}, we propose a novel linearization approach which includes saturation recovery by joint processing of receiver array outputs and reduced complexity IMD compensation after MIMO combining. The goals in this work are to:
\begin{itemize}
\item Design DSP block $g(.)$ to compensate distortion due to saturation of LNA.
\item Design beam-space (post-MIMO-combining) DSP block $f(.)$ to compensate distortion from small scale nonlinearity with reduced complexity as compared to existing solutions.
\end{itemize}

%
%
\section{DSP Algorithms for LNA linearization}
\label{sec:algorithm}
In this section, we present the proposed DSP linearization algorithm. We start with the saturation recovery approach, i.e., mitigation of large scale nonlinearity, and then present the beam space IMD compensation. For reference, we briefly introduce the existing per-antenna linearization approach at the end followed by complexity analysis.

\subsection{Null-space assisted saturation recovery}
\label{sec:clipping}
This subsection focuses on the saturation compensation $g(\mathbf{y}_{\text{ADC}}) = \mathbf{y}_{\text{ADC}} - \mathbf{n}_{\text{sat}}$ by estimating $\mathbf{n}_{\text{sat}}$. We assume the nonlinearity model of LNA does not contain the third-order polynomial.

In single-antenna scenario, the large scale nonlinearity cannot be effectively mitigated since saturation is a non-invertible function. However, when receiver is equipped with large antenna array, the mutual correlation among received signals in $\mathbf{y}_{\text{ADC}}$ can be used. Particularly, our design exploits the fact that the original received signal $\mathbf{x}$ is a linear combination of columns of $\mathbf{H}$ in (\ref{eq:basic_model}), and therefore has zero power in its null-space, i.e., $\mathbf{H}_{\text{null}} \in \mathbb{C}^{\Nr \times\Nr-U }$ such that $\mathbf{H}^{\hermitian}_{\text{null}}\mathbf{H}=\mathbf{0}$. We exploit this property and propose the saturation recovery criterion based on the following minimization of the power of received signal in $\mathbf{H}_{\text{null}}$, 
\begin{align}
\begin{split}
\argmin_{\mathbf{n}_{\text{sat}}} \quad & \|\mathbf{H}_{\text{null}}(\mathbf{y}_{\text{ADC}}-\mathbf{n}_{\text{sat}})\|^2\\
\text{subject to}\quad & [\mathbf{n}_{\text{sat}}]_{\mathcal{S}_\text{n}} = \mathbf{0}
\end{split}
\label{eq:null_space_criterion}
\end{align}
where set $\mathcal{S}_\text{n}$ contains the index of LNA in antenna array, thus RF-chain, that does not have saturated samples and $[\mathbf{n}]_{\mathcal{S}}$ picks the element of $\mathbf{n}$ according to elements in set $\mathcal{S}$. The proposed criterion is most suitable for digital saturation mitigation when $|\mathcal{S}_n|$ is close to $N_{\text{R}}$, i.e., only few amplifiers are saturated. Note that the set $\mathcal{S}_\text{n}$ varies for different samples. Practically, the perfect knowledge of $\mathcal{S}_\text{n}$ is not known to the system. However, set $\mathcal{S}_\text{n}$ can be estimated by using the ADC quantization values. Intuitively, when saturation occurs, the magnitude of the quantized sample is likely to exceed certain threshold $\gamma$. Therefore we propose to use the following method to estimate $\mathcal{S}_\text{n}$ as $
\hat{\mathcal{S}}_{\text{n}} \triangleq\{k:|\mathbf{y}_{\text{ADC,k}}|< \gamma\}.$
In order to capture the fact that $\mathbf{n}_{\text{sat}}$ is a small perturbation due to saturation, we introduce the regularization parameter $\kappa$. Then, the regularized-least-squares can be solved by $\mathbf{p} = \mathbf{H}_{\text{null}}\mathbf{y}_{\text{ADC}}$ where $\mathbf{p}\in\mathbb{C}^{\Nr-U}$ is the instantaneous power in signal null space due to ADC and LNA distortions. Denote $\mathbf{V}=[\mathbf{H}_{\text{null}}]_{\mathcal{S}_c}$, where $\mathcal{S}_{\text{c}} = \{1,\cdots,\Nr\} \backslash \mathcal{S}_{\text{n}}$ is the set that contains elements that are likely to have saturation, and $[\mathbf{A}]_{\mathcal{S}}$ picks rows of $\mathbf{A}$ according to elements in set $\mathcal{S}$.  The solution to minimize power in signal null-space (\ref{eq:null_space_criterion}) is given by
\begin{align}
\begin{split}
\hat{\mathbf{n}}_{\text{sat}} = \left[\mathbf{V}^{\hermitian}\mathbf{V}+\kappa \mathbf{I}\right]^{-1}\mathbf{V}^{\hermitian} \mathbf{p}. 
\end{split}
\label{eq:clipping_est}
\end{align}

\subsection{Beam-space IMD compensation}
\begin{figure}
\begin{center}
\includegraphics[width=0.5\textwidth]{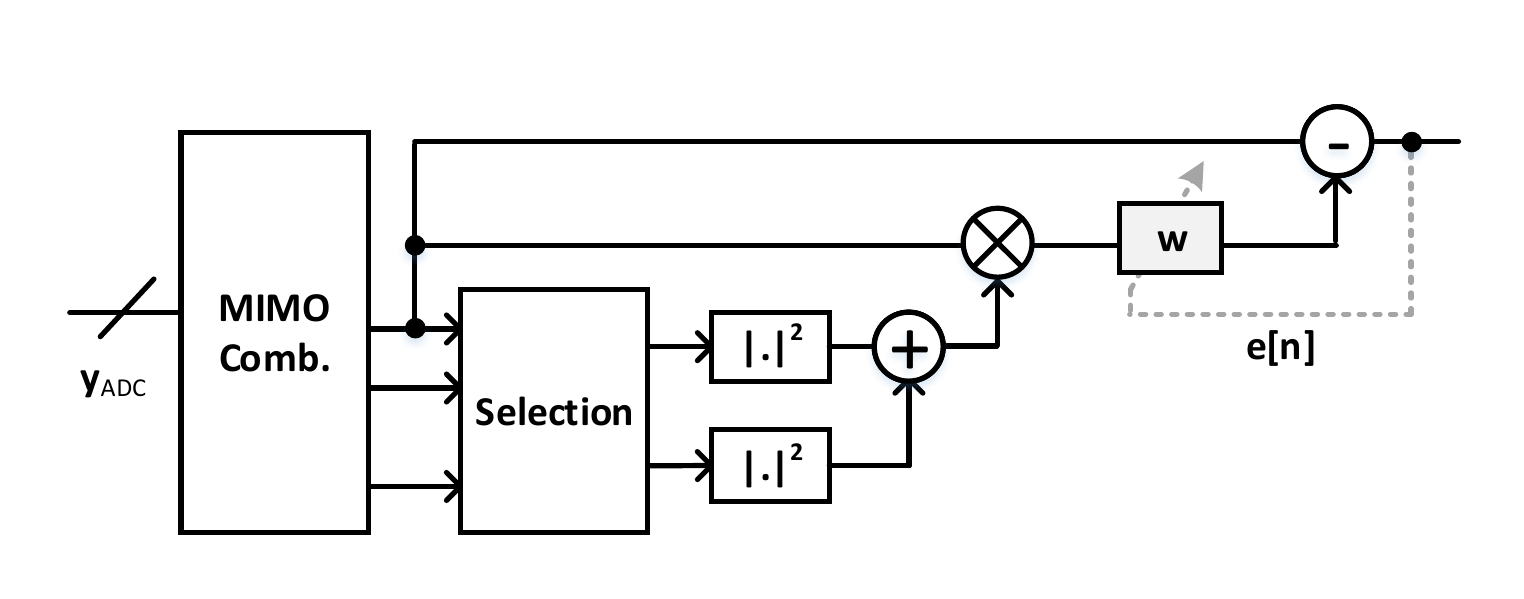}
\end{center}
\vspace{-4mm}
\caption{Proposed processing for beam space linearization.}
\vspace{-4mm}
\label{fig:beam_space_comp}
\end{figure}

In this subsection, we introduce a reduced complexity IMD compensation algorithm. The intuition behind the proposed design can be explained using the following example where effective SIMO channel from $U$ users are orthogonal, i.e., $\mathbf{H}$ is orthonormal. In this special case, the $k$-th post-MIMO-combining symbol can be written as the following expression
\begin{align}
\begin{split}
\hat{s}_k = &\beta_1 s_k + \beta_3 s_k\sum_{i=1}^{U}|\hat{s}_i|^2 \\
&+\underbrace{\beta_3 \sum_{l=1}^{U}\sum_{m=1}^{U}\sum_{n=1}^{U}s_l\hat{s}_ms^*_n\frac{1-e^{j\pi \Nr \sin(\omega_{l,m,n})}}{1-e^{j\pi \sin(\omega_{l,m,n})}}}_{\text{filtered terms}}
\end{split}
\end{align}
where $\omega_{l,m,n} \triangleq \sin^{-1}[\sin(\theta_l)+\sin(\theta_m)-\sin(\theta_n)]$ for clarity. The bracketed filtered term has small power as compared to the rest. The proposed design applies to distortion model approximation $\hat{s}_k = \beta_1 s_k + \beta_3 s_k\sum_{i=1}^{U}|s_i|^2$ and it intends to use $\hat{s}_k\sum_{u=1}^{U}|\hat{s}_u|$ to compensate nonlinear term in $\hat{s}_k$. Due to the difference between $s_u$ and $\hat{s}_u$ for all $u$, a weight $w$ needs to be assigned as presented in Figure~\ref{fig:beam_space_comp} (the example of compensating IMD in the $k=1$, i.e., first stream $\hat{s}_1$). We propose the least-mean-square (LMS) adaptive filter design based on the following criterion
$w_i = \argmin_{w_i} \mathbb{E}\|e_i \|^2.$
Specifically, the error in LMS filter is expressed as $e_i = \hat{s}_i - w_i\hat{s}_i\sum_{u \in \mathcal{U}}|\hat{s}_u|^2$. In the real time adaptation, denote time index as $n$, the update for the $n$-th iteration is expressed as
\begin{align}
w_i[n+1] = w_i[n] + \mu \left(\hat{s}_i[n]\sum_{u\in \mathcal{U}}|\hat{s}_u[n]|^2\right)e^*[n].
\end{align} Apparently, the ideal set $\mathcal{U}$ is $\mathcal{U} = \{1,2,\cdots,U\}$. To reduce the complexity, system can choose only the stream with significant power, i.e.,
 $\mathcal{U} = \{u: \sigma_u^2 \geq \eta \}$.

\subsection{State-of-the-art: per-antenna IMD compensation}
The existing solutions for multiple antenna receiver linearization treat the output of each RF-chain separately in distortion compensation. In other words, the nonlinear compensation is implemented via inverse-function of LNA model (\ref{eq:LNA_NL}), i.e.,
$z_{\text{comp},n} = p^{-1}(y_{\text{ADC,n}}).$
Note that only the small scale function is invertible and therefore large scale is omitted. The practical implementation approaches includes look-up table as well as adaptive filtering.

\subsection{Computational complexity}
This subsection analyzes the computational complexity of the proposed and conventional post-compensation algorithms. For the saturation recovery algorithm (\ref{eq:clipping_est}), the matrix $[\mathbf{V}^{\hermitian}\mathbf{V}+\kappa\mathbf{I}]^{-1}$ cannot be pre-computed in general since matrix $\mathbf{V}$ depends on set $\mathcal{S}_{\text{c}}$ and it varies for different sample indices. Therefore the matrix inversion requires $O(|\Nr-U|^3)$ complexity in terms of multiplication of complex numbers. However, when saturation only occasionally occurs, e.g., $|\mathcal{S}_{\text{c}}| = 1$, matrix inversion lemma can be used to reduce complexity in (\ref{eq:clipping_est}) by $\left[\mathbf{V}^{\hermitian}\mathbf{V}+\kappa \mathbf{I}\right]^{-1}\mathbf{V}^{\hermitian} = 1/(\kappa+\|\mathbf{V}\|^2)\mathbf{V}^{\hermitian}$. In terms of compensating IMD due to the weak nonlinearity, the proposed approach requires $O(|\mathcal{U}|)$ complexity for IMD compensation of a stream when weights converge. The conventional approach requires $O(\Nr)$ for its per antenna operation. Therefore the proposed is more beneficial when MSE of just a few streams are compromised by IMD.
\section{Numerical Evaluation}
\label{sec:simulation results}
This section provides numerical study of nonlinearity in mmWave fully digital receiver in a cellular scenario. We start with link budget study of mmWave uplink, which analyzes the practical input power from mmWave handset, and the nonlinearity  region of the state of the art LNA design. The impact of nonlinear distortion and the efficacy of the proposed digital mitigation is then evaluated.  
\subsection{Case study example}

\begin{table}
\caption{Link Budget Study}
\centering
\begin{tabular}{|l|l|l|l|}
\hline 
\textbf{Case} & 1 & 2 & 3 \tabularnewline
\hline
Frequency (GHz) & 28 & 73 & 140 \tabularnewline
\hline 
Tx Power (dBm) & 23 & 23 & 23\tabularnewline
\hline
Tx Gain (dBm) & 12 & 15 & 21\tabularnewline
\hline
Dist. (m) & 5 & 5 & 5\tabularnewline
\hline
Path Loss (dB) & 75 & 83 & 90\tabularnewline
\hline
Rx Element Gain (dBi) & 3 & 3 & 3\tabularnewline
\hline
Rx Pow (dBm) & -37 & -43 & -48\tabularnewline
\hline
Rx Peak Pow (dBm) & -27 & -33 & -38\tabularnewline
\hline
\end{tabular}
\vspace{-1mm}
\label{tab:link_budget}
\end{table}

\begin{figure}
\begin{center}
\includegraphics[width=0.5\textwidth]{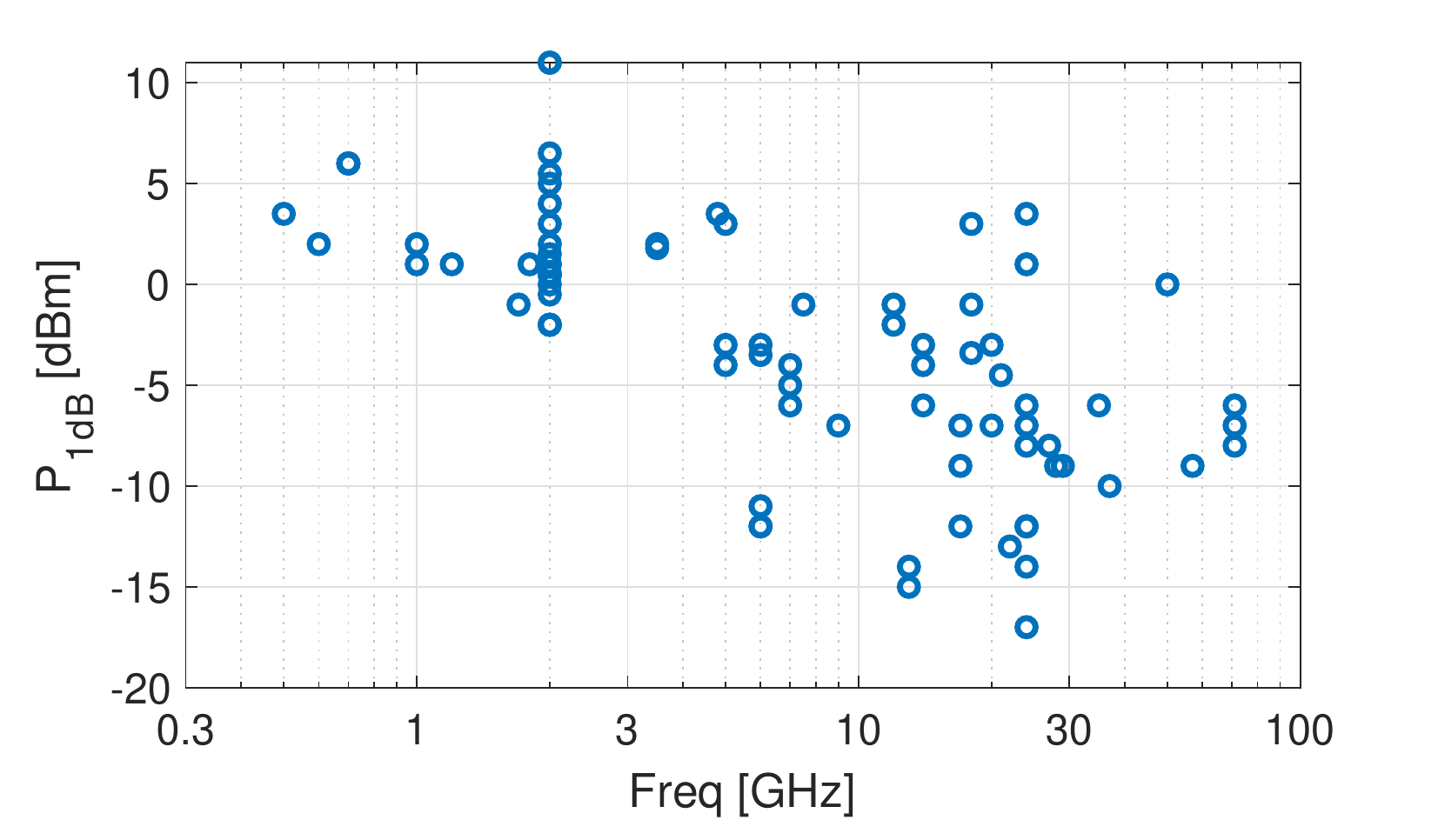}
\end{center}
\vspace{-3mm}
\caption{Survey of the 1dB compression input power with various frequencies. Data source: commercially available LNAs by Analog Devices \cite{AD_survey}.}
\vspace{-5mm}
\label{fig:survey}
\end{figure}

The link budget is shown in Table~\ref{tab:link_budget}. Three mmWave bands anticipated for beyond-5G outdoor use are included, i.e., 28 GHz, 73 GHz, and 140 GHz. The transmitter array gains that correspond to 16, 32, and 128 elements are considered for typical handset dimension. To study the strong nonlinearity regime, the users are assumed to be within 5 meters, and the pathloss is based on recent measurement campaign \cite{8647921}. In addition, although the LNA linearity specifications in above 100 GHz bands are not available, our survey shows a trend of degrading linearity with increase in carrier frequency for commercial products, as shown in Figure~\ref{fig:survey}. The link budget analysis shows that when a handset is placed close enough to the basestation without power control, the receiver will suffer from nonlinearity.

\subsection{Simulation results of nonlinearity mitigation}
We study an UL network setting when the basestation receiver experiences a high dynamic range received power from multiple users. Specifically, 7 out of 8 users are placed close to BS, which results in detrimentally strong input power at the receiver array. Their total power level at the receiver, after considering transmitter array gain, varies from -49 to -40 dBm. The signal of interest has a  power of -70 dBm at each receiver antenna. The ADC model (\ref{eq:ADC_quantization}) assumes uniform quantization, and its overload region is optimally chosen by varying ADC loading fraction \cite{Gersho:1991:VQS:128857}. The normalized distortion $\bar{D}$ of the cascaded system illustrated in Figure~\ref{fig:system_model} is evaluated using \textit{Bussgang linearization} approach \cite{UCSB_LNA_ADC} given original waveform $s$ and estimated $\hat{s}$ after MIMO detection and post-compensation, i.e., 
\begin{align*}
    D = \min_{\alpha} \mathbb{E}\|s- \alpha \hat{s}\|^2,\quad 
    \bar{D} = D/\mathbb{E}\|s\|^2
\end{align*} The receiver thermal noise is omitted since this study focuses on very high SNR regime. The remaining simulation parameters are summarized in Table~\ref{tab:simulation_setting}.


\begin{table}
\caption{Summary of Simulation settings}
\centering
\begin{tabular}{|l|l|}
\hline 
\textbf{Parameter} & \textbf{Value or Setting}\tabularnewline
\hline  
\hline
Rx Array Size& $\Nr = 64$ ULA \tabularnewline
\hline
UE number & $U=8$\tabularnewline
\hline
Channel & LOS; Effective SIMO$^1$ for each user \tabularnewline
\hline
AoA of UEs & Between $-\pi/3$ to $\pi/3$\tabularnewline
\hline
LNA Gain & 15 dB ($\beta_1 = 5.6$)\tabularnewline
\hline
LNA IIP3 & -40 to -16 dBm ($10^3<|\beta_3|<10^6$)\tabularnewline
\hline
ADC Bits & $B$ from 3 to 6\tabularnewline
\hline
ADC loading fraction & Manually tuning for best performance \tabularnewline
\hline
Test Waveform & 16-QAM (5 times oversampling)\tabularnewline
\hline
\multicolumn{2}{l}{1. User steering gain is included in the input power of BS.}
\end{tabular}
\vspace{-1mm}
\label{tab:simulation_setting}
\end{table}

\begin{figure}
\begin{center}
\includegraphics[width=0.5\textwidth]{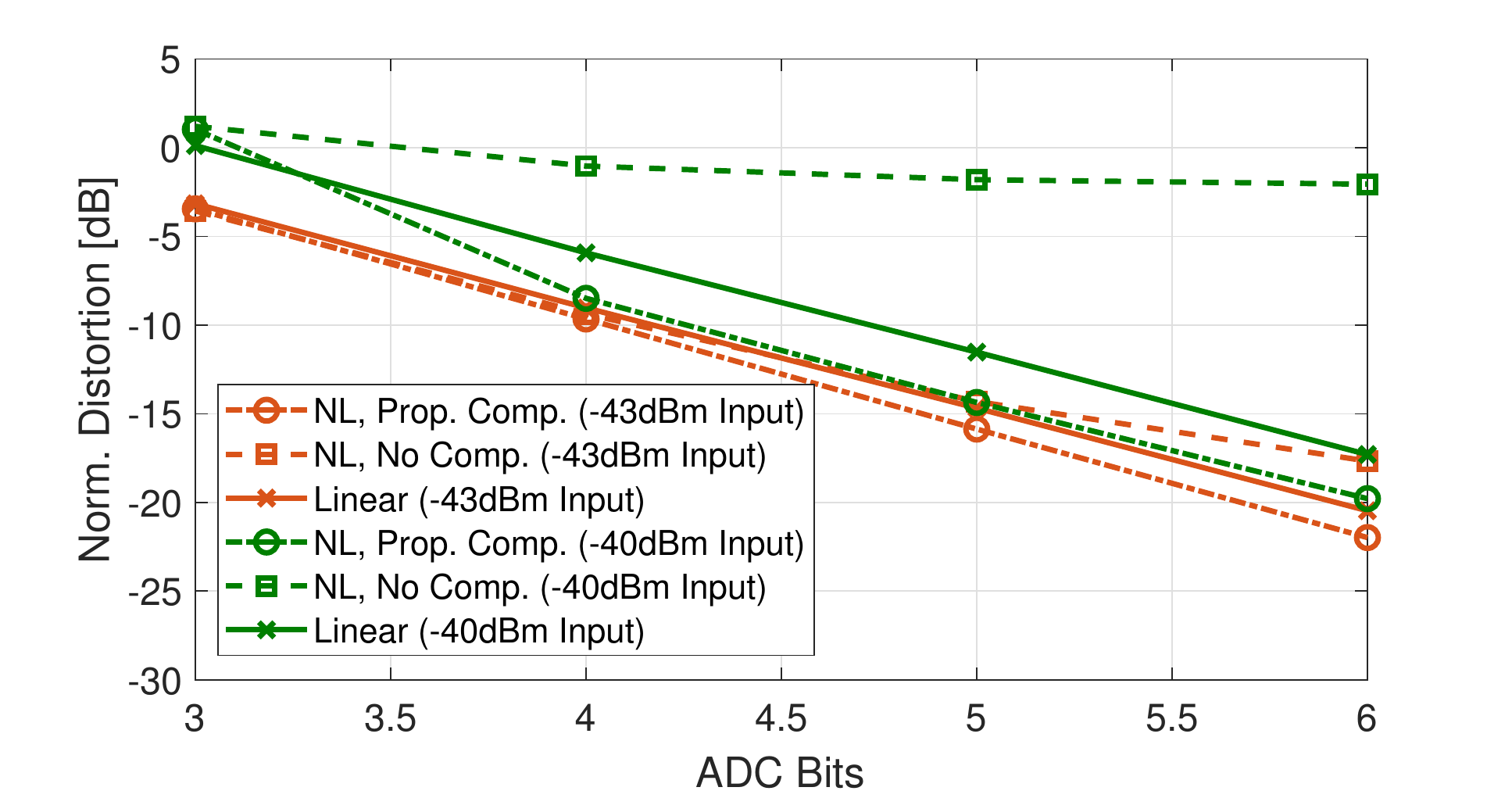}
\end{center}
\vspace{-3mm}
\caption{The distortion power in the stream of interest after MIMO combiner and digital mitigation when LNA saturation is modeled.}
\vspace{-3mm}
\label{fig:distortion}
\end{figure}

In Fig.~\ref{fig:distortion}, we show the distortion power in the stream of interest when LNA is in saturation region. The parameters $\gamma = 0.95V_{\text{max}}$ and $\kappa=0.01$ in Section~\ref{sec:clipping} are heuristically chosen. We have the following observations. Firstly, saturation can severely affect post MIMO combining distortion when input signal is detrimentally strong. Referring to Table~\ref{tab:link_budget}, users within 5 meter distance, i.e., -48 dBm input power of each UE, can severely affect performance of decoding signal from another user far away, when the LNA is assumed to have -30 dBm IIP3. Using the proposed approach, the distortion can be effectively compensated. Secondly, the effectiveness of nonlinearity compensation also depends on the ADC quantization, and generally improves with increased number of bits. 

\begin{figure}
\begin{center}
\includegraphics[width=0.5\textwidth]{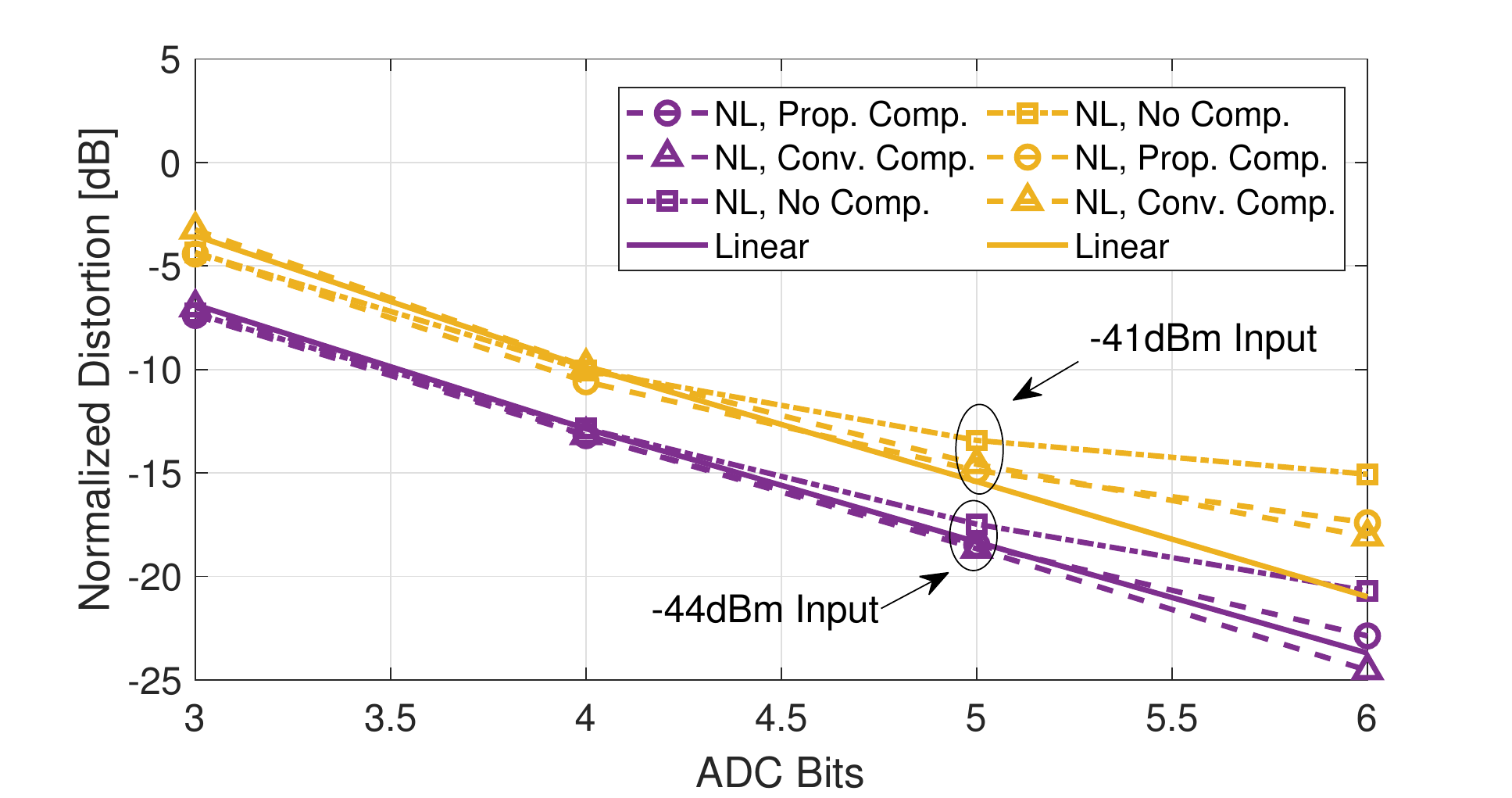}
\end{center}
\vspace{-3mm}
\caption{The distortion power in the stream of interest after MIMO combiner and digital mitigation when LNA polynomial nonlinearity is modeled.}
\vspace{-3mm}
\label{fig:distortion2}
\end{figure}

In Fig.~\ref{fig:distortion2}, the distortion power in the stream of interest is evaluated when LNA has only 3rd order polynomial behavior that corresponds to -30 dBm IIP3. The input power regime when IMD affects system distortion performance is lower than saturation since weak nonlinearity behavior is more sensitive. In the evaluation, the total input power is dominated by a single UE. In this scenario, the proposed algorithm effectively compensates IMD when input is below -41 dBm. The performance of the proposed approach is close to conventional (Conv.) method with per-antenna compensation, but it requires much lower computational complexity.

\begin{figure}
\begin{center}
\includegraphics[width=0.5\textwidth]{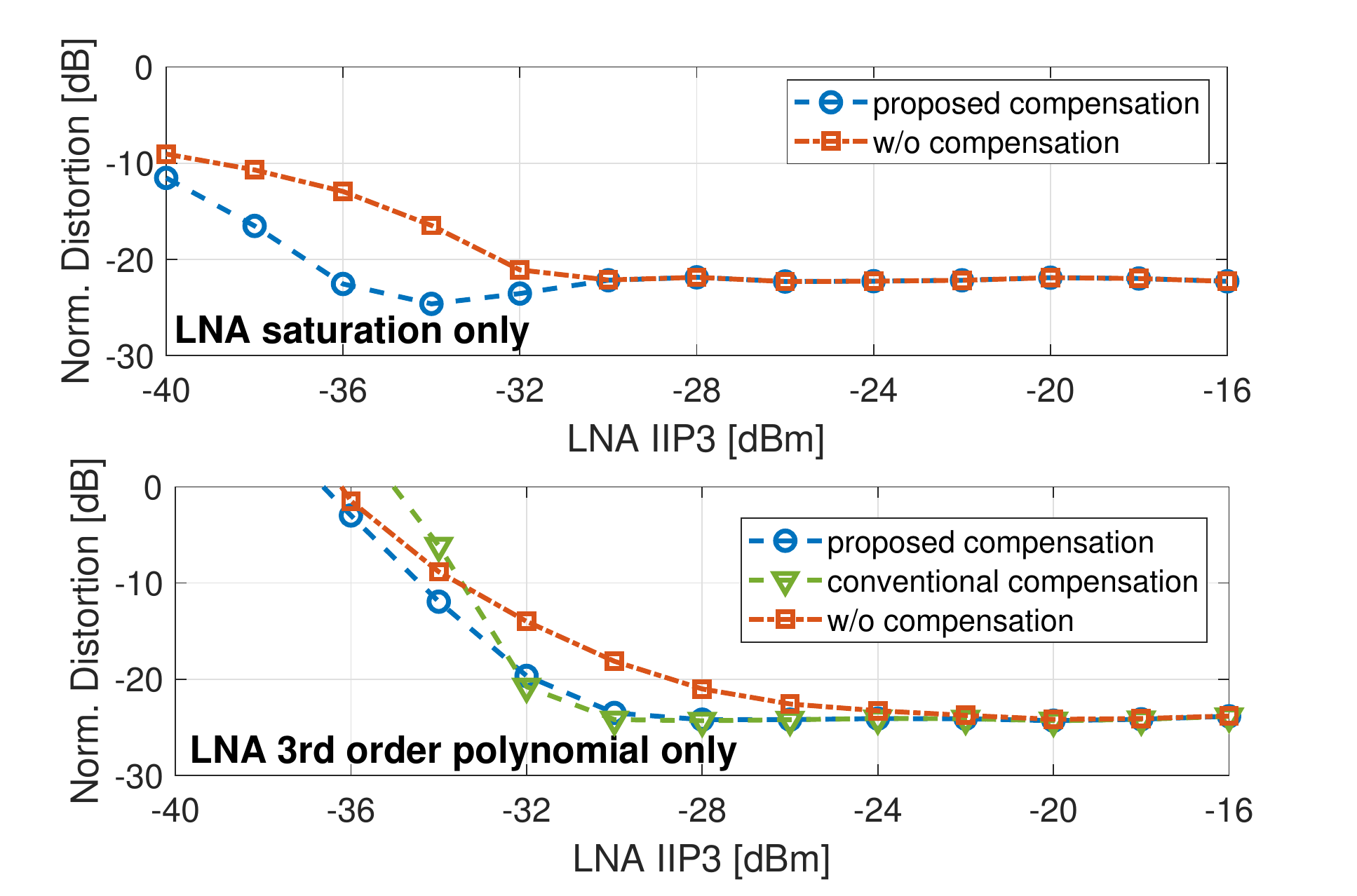}
\end{center}
\vspace{-3mm}
\caption{Normalized power of distortion as function of LNA IIP3. The total input power at BS is -43 dBm and ADC quantization is $B=6$ bits.}
\vspace{-3mm}
\label{fig:distortion3}
\end{figure}

In Fig.~\ref{fig:distortion3}, the relaxation of LNA linearity specification is evaluated. Distortion levels that correspond to ADC quantization and two LNA non-ideal behaviors are evaluated separately. The proposed approach effectively handles saturation with IIP3 within range of [-36, -30] dBm, where a number of antennas experience saturation. In fact, the proposed algorithm recovers those samples in quantization free manner, and therefore the distortion is slightly better than in a scenario when IIP3 is greater than -30 dBm, where none of the amplifiers are saturated. The proposed approach relaxes IIP3 by 9dB by handling either the saturation or the 3rd order polynomial distortion.

%
%
\section{Conclusion and Future Works}
\label{sec:Conclusion}
This work studies the algorithms to mitigate distortion from LNA nonlinearity in fully digital receiver arrays. We have proposed two algorithms. The first approach utilizes the null space of multiuser CSI to recover samples due to amplifier saturation by joint processing of samples from large antenna array. The second approach compensates the IMD from LNA polynomial nonlinearity. As compared to per-antenna mitigation, the proposed approach has less complexity while achieving comparable performance. The evaluation in MU-MIMO uplink reveals that using the proposed approach the required IIP3 of LNA can be relaxed by 9 dB.

As future work, our goal is to improve the proposed algorithm to compensate distortion when LNA has both weak nonlinearity and saturation. In addition, the proposed polynomial nonlinearity compensation will be extended for more complicated channel models. Further, we will investigate the trade-off between relaxed LNA power consumption and increased DSP burden.

\section{Acknowledgemenet}
The authors would like to thank Mohammed Abdelghany (UCSB) and Ali A. Farid (UCSB) for the helpful discussions.


%
%



%
\bibliographystyle{IEEEtran}
\bibliography{IEEEabrv,references}

\end{document}